\documentclass[a4paper,12pt]{article}
\usepackage{epsfig}
\usepackage{graphics}
\def\newpic#1{}

\begin{document}

\begin{center}

{\large \bf ASYMPTOTICS OF THE DEUTERON FORM
FACTORS IN THE NUCLEON MODEL AND JLab EXPERIMENTS}

\end{center}

\begin{center}
A.F. Krutov \footnote{E-mail: krutov@ssu samara.ru}

\textit{Samara State University, 443011 Samara, Russia}

\vspace{1cm}
 V.E. Troitsky
\footnote{E-mail: troitsky@theory.sinp.msu.ru}

\textit{D.V.~Skobeltsyn Institute of Nuclear Physics, Moscow State
University, 119992 Moscow, Russia}

\vspace{1cm}

N.A. Tsirova \footnote{E-mail: ntsirova@ssu samara.ru}

\textit{Samara State University, 443011 Samara, Russia}
\end{center}

\begin{abstract}
Using the  instant form dynamics of Poincar\'e invariant quantum
mechanics and the modified relativistic impulse approximation
proposed previously we calculate asymptotics of electromagnetic
form factors for the deuteron considered as two--nucleon system.
We show that today experiment on the elastic $ed$-scattering has
reached asymptotic regime. The possible range of momentum
transfer when the quark degrees of freedom could be seen in
future JLab experiments is estimated. The explicit relation
between the behavior of deuteron wave function at $r=0$ and the
form factors asymptotics is obtained. The conditions on wave
functions to give the asymptotics predicted by QCD and quark
counting rules are formulated.
\end{abstract}

\vspace{5mm}

\vspace{5mm}

\newpage

Recent experiments with electron accelerators and, especially,
the Jefferson Laboratory (JLab) experiments using continuous
electron beam arouse the interest in the theoretical study of the
hadron electromagnetic structure (see, e.g., \cite{AlK01, MeS04,
Tom05, AmD03, Kar04, HJD05, SeP04, BuB03, PhW05} and the review
\cite{GiG02}). One of the most impressing JLab recent results is
the discovery, in the experiments on polarization transfer in
electron--proton scattering, of a new ("non--Rosenbluth") behavior
of proton form factors, that is of the faster decrease of proton
charge form factor as compared with the magnetic form factor
\cite{Gay02}. The other one is the measurement of the deuteron
tensor polarization component
 $T_{20}(Q^2)$
 ($Q^2=-q^2,\, q$ is the transferred momentum)
in polarization experiments on elastic
electron--deuteron scattering at
$Q^2\simeq$ 2 (GeV/c)$^2$ \cite{AbA00}.
In \cite{KrT07} we have shown that the existing
data for
 $T_{20}(Q^2)$
provide a crucial test for deuteron wave
functions.

The JLab program of investigations on elastic
electron--deuteron scattering at
$Q^2\simeq$ 10 (GeV/c)$^2$ \cite{ArH05}
attracts exclusive attention. Those experiments
will be an important source of information on
nucleon--nucleon interaction at short distances,
on the role of relativistic effects, on exchange
meson (two--particle) currents and also on the
quark degrees of freedom in the deuteron
electromagnetic structure. There exists a hope
that these JLab experiments will help to determine
the limits of the two--nucleon model and to
clarify the interplay between nucleon--nucleon
and quark approaches to the deuteron.
The momentum transfer range
$Q^2\simeq$ 10 (GeV/c)$^2$ (as we show below)
is asymptotical for the deuteron considered
as a nucleon--nucleon system. That is why these
experiments are of great interest. In fact,
 first, in the asymptotical domain there exists
the most surely established QCD prediction for deuteron form
factors \cite{BrF73} (see the quark counting rule prediction, too
\cite{MaM73}). Second, the asymptotic behavior of the nucleon
phenomenological model probably can give a possibility of correct
taking into account of quark degrees of freedom in the framework
of nucleon--nucleon dynamics at short distances.

The present paper is devoted to a theoretical
investigation of asymptotic behavior of the
deuteron form factors at large momentum transfer
in the framework of the nucleon model of
deuteron. We use the relativistic invariant
impulse approximation in a variant of instant
form dynamics of Poincar\'e invariant quantum
mechanics (PIQM) developed in our papers
previously \cite{KrT03-2,KrT02,KrT03,KrT05} and
the results of our paper \cite{KrT07-2} where a
theorem on asymptotic estimation of the
quantities in question is proved which is valid
in relativistic as well as in nonrelativistic
case.

Let us consider first the asymptotics of the deuteron form
factors in the nonrelativistic impulse approximation. The
standard expressions for electromagnetic deuteron form factors can
be written in terms of wave functions in momentum representation
in the following way (see, e.g. \cite{JaM72}):
$$
G^{NR}_C(Q^2) = \sum_{l,l'}\int\,k^2\,dk\,k'\,^2\,dk'\, u_l(k)\,
\tilde g^{ll'}_{0C}(k\,,Q^2\,,k')\, u_{l'}(k')\;,
$$
$$G^{NR}_Q(Q^2) =
\frac{2\,M_d^2}{Q^2}\,\sum_{l,l'}\int\,k^2\,dk\,k'\,^2\,dk'\,
u_l(k)\,\tilde g^{ll'}_{0Q}(k\,,Q^2\,,k')\,u_{l'}(k')\;,
$$
\begin{equation}\label{GqGNIP}
G^{NR}_M(Q^2) =-\,M_d\,\sum_{l,l'}\int\,k^2\,dk\,k'\,^2\,dk'\,
u_l(k)\,\tilde g^{ll'}_{0M}(k\,,Q^2\,,k')\, u_{l'}(k')\;.
\end{equation}
Here $G^{NR}_i\,,\,\,i=C,Q,M$ are charge,
quadrupole and magnetic dipole nonrelativistic
form factors,
$u_l(k)$ are model deuteron wave functions in
momentum representation,
$l,l'=0,2$ are orbital angular momenta, $M_d$ is
the deuteron mass,
$\tilde
g^{ll'}_{0i}\,,\,\,i=C,Q,M$ are nonrelativistic
free two--particle form factors
($2\times 2$ matrices). The
explicit form of $\tilde g^{ll'}_{0i}$, which
is rather cumbersome,
is given in the Appendix of \cite{KrT07}.

In \cite{KrT07-2} the authors, on the base of
a theorem proven there, obtained for the form
factors (\ref{GqGNIP}) the asymptotic
expansions in inverse powers of
$Q^2$ at $Q^2\to\infty$. Two leading terms
have the form:
$$
G_i^{NR}(Q^2) \sim A_i\frac{2\sqrt{\pi}}{r_0^3Q}\left(5
f^{NR}_i(t,Q^2,t'(t))+\frac{1}{r_0^2}\frac{\partial^2}{\partial
t^2}f^{NR}_i(t,Q^2,t'(t)) + \right.
$$
\begin{equation}
\left.\left. +\frac{4\sqrt{2}}{Qr_0^2}\frac{\partial}{\partial
t'}f^{NR}_i(t,Q^2,t'(t))-
\frac{16}{Q^2r_0^2}f^{NR}_i(t,Q^2,t'(t))\right)
\right|_{{{t=0}\atop{t'=0}}}, \label{anywfner}
\end{equation}
with
$$
f^{NR}_i(t,Q^2,t')=\sum\limits_{l,l'=0,2}k^2\;k'\,^{2}
u_l(k)\;\tilde{g}^{ll'}_{0i}(t,Q^2,t')\;u_{l'}(k')\;,
$$
$$
k=\frac{1}{\sqrt{2}}\left(t'+ t\right) + \frac{Q}{4}\;, \quad
k'=\frac{1}{\sqrt{2}}\left(t'- t\right) + \frac{Q}{4}\;,
$$
$i=C,Q,M$, $\;A_C=1,\;A_Q=2M_d^2/Q^2\,,\;A_M=-M_d$,
the function  $t'(t)$ describes the boundary of
the domain of integration in
(\ref{GqGNIP}) (see \cite{KrT07}), the parameter
$r_0$ is defined by the deuteron
matter radius in different models of $NN$-
interaction \cite{KrT07-2}.

In modern calculations the deuteron wave functions
 usually are of the
following analytic form in the momentum representation (see, e.g.,
\cite{LaL81,Mac01,KrT07wf}):
\begin{equation}
u_0(k)=\sqrt{\frac{2}{\pi}}\sum\limits_j
\frac{C_j}{(k^2+m_j^2)},\;\quad
u_2(k)=\sqrt{\frac{2}{\pi}}\sum\limits_j
\frac{D_j}{(k^2+m_j^2)}.\; \label{funcu}
\end{equation}
or in the coordinate representation:
\begin{equation}
u_0(r) = \sum\limits_j {C_j}{\exp\left(-m_j\,r\right)}\;,\quad
u_2(r) = \sum\limits_j {D_j}{\exp\left(-m_j\,r\right)} \left[1 +
\frac{3}{m_j\,r} + \frac{3}{(m_j\,r)^2}\right]\;,
\label{funcukoord}
\end{equation}
here $m_j = \alpha + m_0\,(j-1)\;,\; \alpha =
\sqrt{M\,\left|\varepsilon_d\right|},$$\;M$ is average nucleon
mass, $\varepsilon_d$ -- the binding energy of
the deuteron. The coefficients
$C_j,\,D_j$, the maximal value of the index $j$
and $m_0$ are determined by the best
fit of the corresponding solution of the
Schr\"odinger equation.

The standard behavior of the functions at short
distances:
\begin{equation}
u_0(r)\;\sim\;r\;,\quad u_2(r)\;\sim\;r^3\;, \label{atorigin}
\end{equation}
is provided by imposing the following conditions
on the coefficients in
(\ref{funcukoord}):
\begin{equation}
\sum\limits_j {C_j} = 0\;,\quad \sum\limits_j {D_j} =
\sum\limits_j {D_j}{m_j^2} = \sum\limits_j \frac{D_j}{m_j^2} =
0\;. \label{cond}
\end{equation}
Using the explicit form of $\tilde
g^{ll'}_{0i}\,,\,\,i=C,Q,M$ from \cite{KrT07}
and taking into account the Eq.
(\ref{cond}), we obtain from
(\ref{anywfner}), (\ref{funcu}) the main
asymptotic terms of the nonrelativistic
deuteron form factors:
\begin{equation}
G_C^{NR}\sim
\frac{1}{Q^8}\frac{2^{15}}{\sqrt{\pi}r_0^3}\left[\sum\limits_{j}C_{j}m_j^2\right]^2
\left(G_E^p(Q^2)+G_E^n(Q^2)\right)\;, \label{gnrc}
\end{equation}

\begin{equation}
G_Q^{NR}\sim
3\,M_d^2\frac{1}{Q^{12}}\frac{2^{{41}/{2}}}{\sqrt{\pi}r_0^3}
\left[\sum\limits_{j}C_{j}m_j^2\right]\left[\sum\limits_{j}D_{j}m_j^4\right]
\left(G_E^p(Q^2)+G_E^n(Q^2)\right)\;, \label{gnrq}
\end{equation}

\begin{equation}
G_M^{NR}\sim
\frac{1}{Q^8}\frac{2^{15}M_d}{\sqrt{\pi}r_0^3M}\left[\sum\limits_{j}C_{j}m_j^2\right]^2
\left(G_M^p(Q^2)+G_M^n(Q^2)\right)\;, \label{gnrm}
\end{equation}
here $G_{E(M)}^{p(n)}$ are charge
and magnetic form factors of proton
and neutron, correspondingly.

Let us note that the asymptotic forms
(\ref{gnrc})-(\ref{gnrm}) are valid for any model
of $NN$-interaction if
(\ref{atorigin}) is satisfied.
The fact that
$G_Q^{NR}$ decreases faster than other form
factors is due to the faster decreasing of the
$D$-wave function at $r\to 0$ as compared with
$S$-wave function (\ref{atorigin}). From the
mathematical point of view the form of the leading
terms in  (\ref{gnrc})-(\ref{gnrm}) is the
consequence of the conditions
(\ref{cond}). Any modification of
these conditions (or, equivalently,
of the conditions (\ref{atorigin}))
changes (\ref{gnrc})-(\ref{gnrm}) as well.

The asymptotic expansions of the deuteron form
factors contain the nucleon form factors. We use
the standard dipole fit for magnetic nucleon form
factors \cite{GiG02}:
\begin{equation}
\frac{G_M^p(Q^2)}{\mu_p}=\frac{G_M^n(Q^2)}{\mu_n} =
G_d(Q^2)\;,\,\,\,
G_d(Q^2)=\left(1+\frac{Q^2}{a_d^2}\right)^{-2}\;, \label{GN}
\end{equation}
with $a_d^2=$0.71 GeV$^2$. Here $\mu_p,\;\mu_n\;$
-- are proton and neutron magnetic
moments, corespondingly.
As the neutron charge form factor is small enough
we neglect its contribution. In this connection,
one has to notice that its behavior at large
momentum transfer has also the dipole form
(see, e.g., \cite{KrT03-2}), so that its
contribution would not change our results.
We assume the "non-Rosenbluth"
\cite{Gay02} behavior for the proton charge form
factor and so we obtain the faster decrease of
the deuteron charge form factor as compared with
the magnetic one. Let us note that the
asymptotics of nucleon form factors
(\ref{GN}) ($\sim Q^{-4}$) agrees with the
predictions of QCD for nucleons
\cite{BrF73}.

Usually describing the deuteron one is dealing
with the asymptotic behavior of the function
(see, e.g.,\cite{ArH05}):
\begin{equation}
F_d(Q^2)=\sqrt{A(Q^2)}\;, \label{Fd}
\end{equation}
where the structure function $A(Q^2)$,
which enters the electron--deuteron scattering
cross--section, can be written in terms of the
deuteron form factors in the following way
\cite{GiG02}:
\begin{equation}
A(Q^2) = G_C^2(Q^2) + \frac{8}{9}\eta^2\,G_Q^2(Q^2) +
\frac{2}{3}\eta\,G_M^2(Q^2)\;,\quad \eta = \frac{Q^2}{4M_d^2}\;.
\label{A}
\end{equation}
After substituting the expansions
(\ref{gnrc})-(\ref{gnrm}) in (\ref{A})
one can see that the main contribution to
(\ref{Fd}) or (\ref{A}) comes from the term
containing the deuteron magnetic form factor.
Note, that the "non-Rosenbluth" behavior of
proton form factor reinforces the main role of
this term. Using (\ref{GN}) we obtain the
asymptotics of
$A(Q^2)$ in the nonrelativistic impulse
approximation:
\begin{equation}
A^{NR}(Q^2)\;\sim\;\frac{1}{Q^{22}}\frac{2^{29}(\mu_p+\mu_n)^2
a_d^8 }{3\pi r_0^6 M^2}\left[\sum\limits_{j}C_{j}m_j^2\right]^4.
\label{ANR}
\end{equation}
Let us emphasize that our approach permits us to
write the prefactor in (\ref{ANR}) explicitly.
Other authors dealing with the deuteron
asymptotics (see, e.g., \cite{ArC80}),
usually derive the power of momentum transfer
only. Our form of the power-law dependence is
\begin{equation}
F_d^{NR}(Q^2)\;\sim\;Q^{-11}\; \label{asner}
\end{equation}
in nonrelativistic case.

However, at large momentum transfer it is important to use
relativistic approach. To give the relativistic description of
the deuteron we use here the formalism of the Poincar\'e
invariant quantum mechanics (PIQM). Today this approach is one of
the effective approaches to the structure of composite systems
(see, e.g.,\cite{GiG02}). Actually we use a variant of the
instant form dynamics of PIQM developed by the authors (see
\cite{KrT02,KrT03,KrT05})).In our approach the relativistic
electromagnetic form factors of a composite system can be written
in a form similar to the nonrelativistic case (\ref{GqGNIP}). The
equations obtained in the modified Lorentz-invariant relativistic
impulse approximation are given in the paper \cite{KrT07} and
have the following form:
$$
G^R_C(Q^2) = \sum_{l,l'}\int\,d\sqrt{s}\,d\sqrt{s'}\,
\varphi_l(s)\, g^{ll'}_{0C}(s\,,Q^2\,,s')\, \varphi_{l'}(s')\;,
$$
$$
G^R_Q(Q^2) =
\frac{2\,M_d^2}{Q^2}\,\sum_{l,l'}\int\,d\sqrt{s}\,d\sqrt{s'}\,
\varphi_l(s)\,g^{ll'}_{0Q}(s\,,Q^2\,,s')\,\varphi_{l'}(s')\;,
$$
\begin{equation}
\label{GqGRIP}
G^R_M(Q^2)=-\,M_d\,\sum_{l,l'}\int\,d\sqrt{s}\,d\sqrt{s'}\,
\varphi_l(s)\,g^{ll'}_{0M}(s\,,Q^2\,,s')\, \varphi_{l'}(s')\;.
\end{equation}
Here $\;\varphi_l(s),\;l,l'=0,2$ are the deuteron
wave functions in the sense of PIQM,
$g^{ll'}_{0i}\,,\;i=C,Q,M$ are relativistic free
two-particles charge, quadrupole and magnetic
form factors \cite{KrT07}.

The deuteron wave functions in the sense of PIQM
are solutions of the eigenvalue problem for a
mass squared operator for the deuteron
$\hat
M^2_d\, |\psi\rangle = M^2_d\,|\psi\rangle$ (see,
e.g., \cite{GiG02, KrT02}). This equation
coincides with the nonrelativistic
Schr\"odinger equation within a second order in
deuteron binding energy
${\varepsilon_d^2}/{4M}$ ($M$ is an averaged
nucleon mass). The value of this quantity is
small, so the deuteron wave functions in the
sense of PIQM differ from nonrelativistic wave
functions by the normalization only. In the
relativistic case the wave functions are
normalized with relativistic density of states:
\begin{equation}
\sum_{l=0,2}\,\int_0^\infty\varphi^2_l(k)\,\frac{dk}{2\sqrt{k^2 +
M^2}}=1\;,\quad \varphi_l(k) = \sqrt[4\ ]{s}\,k\,u_l(k)\;,\quad
s=4(k^{2}+M^{2})\;.\label{wfRHD}
\end{equation}
The nonrelativistic formulae
(\ref{GqGNIP}) given in \cite{JaM72},
can be obtained from relativistic ones
(\ref{GqGRIP}) in the nonrelativistic limit.

Two leading terms of the asymptotic expansion of
the deuteron form factors have the form
\cite{KrT07-2} (see also the notes to Eq.
(\ref{anywfner})):
$$
G^R_i(Q^2) \sim A_i\frac{2^{{7}/{2}}\sqrt{\pi M}}{r_0^3Q}
\left(5f^R_i(t,Q^2,t'(t))+\frac{2^{{7}/{2}}}{r_0^2}\frac{\partial}{\partial
t'}f^R_i(t,Q^2,t'(t))  \right. +
$$
\begin{equation}
\left.\left. +\frac{8 M}{r_0^2 Q}\frac{\partial^2}{\partial
t^2}f^R_i(t,Q^2,t'(t)) \right) \right|_{{{t=0}\atop{t'=0}}}\;,
\label{anywfrel}
\end{equation}
where
$$
f^R_i(t,Q^2,t')=\sum\limits_{l,l'=0,2}\frac{Q}{4\sqrt{ss'}}\varphi_l(s)\;g^{ll'}_{0i}(t,Q^2,t')\;
\varphi_{l'}(s')\;,
$$
$$
s=\frac{1}{\sqrt{2}}\left(t' + tQ\right) + 2 M^2 + M\sqrt{Q^2 + 4
M^2}\;,\quad s'=\frac{1}{\sqrt{2}}\left(t' - tQ\right) + 2 M^2 +
M\sqrt{Q^2 + 4 M^2}\;.
$$

Using (\ref{funcu}), (\ref{wfRHD}),
 (\ref{anywfrel}) and the explicit forms of
$g^{ll'}_{0i}\,,\;i=C,Q,M$ from \cite{KrT07}, we obtain the main
terms of relativistic asymptotic expansions (\ref{GqGRIP}),
(\ref{anywfrel}), which enables us to write the relativistic
asymptotics of the function $A(Q^2)$ in (\ref{A}) in the form:
\begin{equation}
A^R(Q^2)\;\sim\;\frac{Q^6}{2^{11}M^6}\,A^{NR}(Q^2)\;. \label{Arel}
\end{equation}
So, the relativistic asymptotics of the form
factor (\ref{Fd}) is:
\begin{equation}
F^R_d(Q^2)\;\sim\;Q^{-8}\;.
\label{asrel}
\end{equation}
The relativistic effects slow down the decreasing
of the form factors as compared with the
nonrelativistic case
(\ref{ANR}),
(\ref{asner}).

The comparison of the first and the second leading terms in the
obtained asymptotic expansions for different models
\cite{LaL81,Mac01,KrT07wf,ShM03} of $NN$-interaction shows that
for all of them the asymptotic form (\ref{Arel}), (\ref{asrel})
is valid at $Q^2\;\simeq 6$ (GeV/c)$^2$. So, the existing JLab
experiments at large momentum transfer \cite{Ale99} have already
reached the asymptotic region predicted by relativistic
two-nucleon  deuteron model.

In this connection it is of interest to compare directly the
obtained asymptotic predictions with experimental data. Fitting
the existing experimental points for seven highest attained values
of momentum transfer \cite{Ale99} in region 3.040--5.955
(GeV/c)$^2$ by a power-law function we obtain the following
estimation for (\ref{Fd}) with $\chi^2=3.93\cdot 10^{-9}$:
\begin{equation}
F_d^{exp}(Q^2)\;\sim\; \frac{1}{\left(Q^2\right)^{3.76\pm0.41}}\;.
\label{ase}
\end{equation}
So, comparing (\ref{asner}), (\ref{asrel}) and (\ref{ase}) one can
see that up to fitting accuracy, the experimental data are
described by the relativistic formula (\ref{asrel}). Let us note
that this result does not depend on the actual model of
$NN$-interaction and in fact is due to general conditions
(\ref{atorigin}) and (\ref{cond}) only. So, in the recent JLab
experiments the range of momentum transfers, which can be
characterized as asymptotic one for our relativistic two--nucleon
model of deuteron, is reached. Let us emphasize that this
concerns our relativistic approach only. For example, in the
relativistic approach of \cite{ArC80} the asymptotic decrease is
faster than the experimental one and even faster than that of the
quark counting prediction.

The numerical comparison of our asymptotic prediction
(\ref{Arel}) (including prefactor) for different model
interactions \cite{LaL81,Mac01,KrT07wf,ShM03} with experimental
values of $A(Q^2)$ demonstrates that our theoretical curves lay
lower than the experimental. The same is true about our full
(without asymptotic expansion) relativistic calculation using Eqs.
(\ref{funcu}), (\ref{A}), (\ref{GqGRIP}), (\ref{wfRHD}). It is
worth to notice that the relativistic effects shift the curve
closer to the experimental one as compared with nonrelativistic
case.

\begin{figure}
\begin{center}
\includegraphics[scale=1]{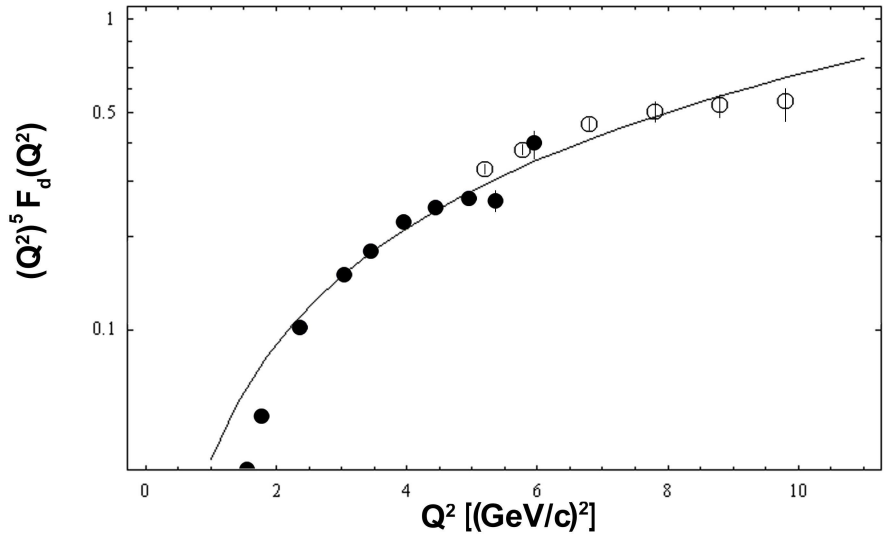}
\caption{Experimental data for the deuteron form factor
(\ref{Fd}). Dark dots are experimental dots obtained in JLab
\cite{ArH05}, light dots are projected in future experiment dots
\cite{ArH05}, solid line is our fit of existing experimental data
at high momentum transfer (\ref{ase})} \label{dff}
\end{center}
\end{figure}

The asymptotics (\ref{ANR}),(\ref{asner})
are obtained in the framework of
the nonrelativistic impulse approximation. The
asymptotics (\ref{Arel}), (\ref{asrel})
correspond to the  Lorentz-invariant modified
impulse approximation \cite{KrT02}. In this
connection it is interesting to estimate the
asymptotic behavior of the contributions of terms
out of impulse approximation, in particular, of
meson exchange currents (MEC). The agreement of
experimental asymptotics of the deuteron form
factors (\ref{ase}) with that obtained in the
relativistic impulse approximation (\ref{asrel})
means that the MEC contribution at large momentum
transfer either has the same power law dependence
on $Q^2$ as in (\ref{asrel}) or decreases faster.
It is possible that this fact is due to a fast
decreasing of transfer mesons form factors (for
example of $\rho\pi\gamma$-form factor) at
$Q^2\to\infty$. This problem will be considered
in detail elsewhere.

Let us compare our results with those of the quark
approach and of QCD. At $Q^2\to\infty$ there
exists the well established prediction in the
framework of these approaches
\cite{BrF73,MaM73}:
\begin{equation}
F_d(Q^2)\;\sim\; Q^{-10}\;. \label{asqcd}
\end{equation}
As one can see, this prediction does not agree with the current
experiment (\ref{ase}). In other words the quark degrees of
freedom are not seen in the range of today momentum transfer. The
hope to see quark degrees of freedom is connected with the
possibility of obtaining in JLab future experiments the deviation
of power law dependence of $A(Q^2)$ and $F_d(Q^2)$ from that
given by our relativistic model (\ref{Arel}), (\ref{asrel}) and
by today experiment (\ref{ase}).

In figure \ref{dff} the projected results
of future JLab experiments are presented
\cite{ArH05}. The saturation is interpreted
\cite{ArH05} as due to quark degrees of freedom.
One can see that this fact may take place
in the range of $Q^2\simeq 10$ (GeV/c)$^2$,
where the projected data leaves our fit
(\ref{ase}).

Let us discuss one more problem connected with the deuteron form
factor asymptotics, namely, the possibility to incorporate the
QCD predictions in the nucleon--nucleon dynamics. One of the
possibilities of obtaining the asymptotics (\ref{asqcd}) in
nucleon physics is to put some conditions on nucleon form factors
in (\ref{gnrc})--(\ref{gnrm}). However, this way can not be
considered as self consistent. More correct way is to use the QCD
prediction ($\sim Q^{-4}$) coinciding at $Q^2 \to\infty$ with the
result of dipole fit (\ref{GN}) used while obtaining the
asymptotic estimations (\ref{asner}), (\ref{asrel}).

So, we formulate the  problem in
the following way: what kind of behavior at small
distances would have the deuteron wave function,
or, in other words, how the nucleon--nucleon
potential at short distances has to be modified
-- in order to obtain the asymptotics of
electromagnetic deuteron form factors predicted
by QCD? The answer can be obtained easily and the
analyzis shows that the quark model asymptotics
could be derived in the nucleon dynamics formalism
if in addition to the conditions
(\ref{cond}) for the $s$-wave function (\ref{funcu})
the following condition is imposed:
\begin{equation}
\sum\limits_j {C_j}{m_j^2} = 0\;. \label{newcond}
\end{equation}

This condition means that in the vicinity of zero
the wave function has the following form:
\begin{equation}
u_0(r) \sim r + a\,r^3\;,\;u_0''(0)=0\;. \label{newbeh}
\end{equation}

So, we have solved some kind of inverse problem:
we found a condition for the deuteron wave
function to give the asymptotic behavior of the
deuteron electromagnetic form factors  (in the
framework of the nucleon model of deuteron) given
by quark approach. The presence of the quark
degrees of freedom changes the deuteron wave
function (\ref{funcu}) following the conditions
(\ref{newcond}), (\ref{newbeh}).

To conclude, in the present paper the following
results are obtained.

The asymptotics of the deuteron form factors at
large momentum transfer is derived in
nonrelativistic as well as in relativistic case
in the framework of impulse approximation in the
PIQM instant form dynamics.

The relativistic effects slow down
the asymptotic decreasing of form factors
and result in the power-law
dependence on  $Q^2$ which coincides with the
experimental data.

It is shown that the range of momentum transfers
of recent JLab experiments may be considered as
asymptotic for relativistic two--nucleon deuteron
model.

By comparison of the calculated asymptotics with experimental
data it is established that quark degrees of freedom are not seen
in today experiment and could be seen in future JLab experiments
when the deviation of experiment from the results obtained in
this paper would be reached.

The condition for the deuteron wave function at
$r=0$ to provide the asymptotic
behavior of deuteron form factors predicted by
QCD is obtained.

This work was supported in part by Russian
Foundation for Basic Researches (grant No
07-02-00962).

\end{document}